\documentclass[preprint,12pt,3p]{elsarticle}

\usepackage{amssymb}
\usepackage{amsmath}
\usepackage{caption}
\usepackage{subcaption}
\usepackage[]{algorithm2e}
\usepackage{multirow}
\usepackage{graphicx}
\usepackage{mathtools}
\usepackage{physics}
\usepackage{hyperref}
\usepackage{physics}

\newcommand{\opt} {OPT }

\DeclareMathOperator*{\minimize}{minimize}
\DeclareMathOperator*{\argmin}{argmin}
\journal{Microprocessors and Microsystems}
\newcommand{\hide}[1] {}
\begin{document}

\begin{frontmatter}

\title{Formal Constraint-based Compilation \\ for Noisy Intermediate-Scale Quantum Systems}

\author[label1]{Prakash Murali\corref{cor1}}
\address[label1]{Princeton University}
\cortext[cor1]{Corresponding author. Full address: Department of Computer Science, Princeton University, 35, Olden Street, Princeton NJ 08540}

\ead{pmurali@princeton.edu}

\author[label2]{Ali Javadi-Abhari}
\address[label2]{IBM Thomas J. Watson Research Center}
\ead{ali.javadi.abhari@gmail.com}

\author[label3]{Frederic T. Chong}
\address[label3]{University of Chicago}
\ead{chong@cs.uchicago.edu }

\author[label1]{Margaret Martonosi}
\ead{mrm@princeton.edu}

\begin{abstract}
Noisy, intermediate-scale quantum (NISQ) systems are expected to have a few hundred qubits, minimal or no error correction, limited connectivity and limits on the number of gates that can be performed within the short coherence window of the machine. The past decade's research on quantum programming languages and compilers is directed towards large systems with thousands of qubits. For near term quantum systems, it is crucial to design tool flows which make efficient use of the hardware resources without sacrificing the ease and portability of a high-level programming environment. In this paper, we present a compiler for the Scaffold quantum programming language in which aggressive optimization specifically targets   NISQ machines with hundreds of qubits. Our compiler extracts gates from a Scaffold program, and formulates a constrained optimization problem which considers both program characteristics and machine constraints. Using the Z3 SMT solver, the compiler maps program qubits to hardware qubits, schedules gates, and inserts CNOT routing operations while optimizing the overall execution time. The output of the optimization is used to produce target code in the OpenQASM language, which can be executed on existing quantum hardware such as the 16-qubit IBM machine. Using real and synthetic benchmarks, we show that it is feasible to synthesize near-optimal compiled code for current and small NISQ systems. For large programs and machine sizes, the SMT optimization approach can be used to synthesize compiled code that is guaranteed to finish within the coherence window of the machine.
\end{abstract}


\end{frontmatter}

\section{Introduction}
The promise of quantum computing (QC) is to provide the hardware and software environment for tackling classically-intractable problems. The fundamental building block of a quantum computer is a qubit or  quantum bit. In the circuit model of quantum computation, quantum programs can be viewed as a series of operations (gates) applied on a set of qubits. These gates may act on a single qubit or on states constructed using multiple qubits. 

Building operational quantum computers requires overcoming significant implementation challenges.  For useful quantum computation, the state of a qubit should be coherent for a long duration of time, the error rates of gates should be low and unwanted quantum interactions should be minimized.

QC's hardware challenges have been partially overcome on small scales (5-20 qubits) using Nuclear Magnetic Resonance \cite{nmr1, nmr2}, trapped ions \cite{trappedion1, trappedion2}, and superconducting qubits \cite{superc1, superc2} among others. Current systems using these technologies have limited coherence time (few hundred microseconds for superconducting qubits), noisy operations (error rates close to 0.01), and limited qubit connectivity. As implementation techniques improve,  these Noisy Intermediate Scale Quantum (NISQ) systems are expected to scale to a few hundred qubits, still with minimal or no error correction and limited connectivity. NISQ systems will also have limited coherence time, allowing at most a few thousand gates to be executed \cite{nisq}.

The last decade's research on quantum computing and compilers has focused on methods for reliable fault tolerant computation using large machines with thousands of qubits \cite{survey1, Devitt2013, ali}. Under tight NISQ constraints, however, it is crucial to design tool flows which make efficient use of the limited hardware resources without sacrificing the ease and portability of a high-level programming environment. In this vein, this paper describes and evaluates a compiler for programs written in a high-level language targeted for NISQ machines with hundreds of qubits. 

Our compiler takes as input a QC program written in the Scaffold language.  The Scaffold language is a QC extension of C. Scaffold features automated gate decomposition and quantum logic synthesis from classical operations. Scaffold programs are independent of the size, qubit technology, connectivity, and error characteristics of the machine. These features provide portability and allow users to express their algorithms at a high level---in terms of logic operations and quantum functions, rather than a more circuit-oriented gate-level description of the intended computation. 

To compile Scaffold programs for NISQ systems, we use an optimization based approach. We express the compilation problem as a constrained optimization problem which incorporates both program and machine characteristics. Using the Z3 Satisfiability Modulo Theory (SMT) solver \cite{z3}, the compiler maps program qubits to hardware qubits, schedules gates, and inserts CNOT routing operations while optimizing the overall execution time. The output of the solver is a near-optimal spatiotemporal mapping that is used to produce target code in the OpenQASM language. The target code can be directly executed on existing quantum hardware such as the 16-qubit machine from IBM. To scale our method to large qubit and gate count, we have developed a heuristic approach which also uses the SMT solver.

Our experiments demonstrate the optimal compilation of programs for the Bernstein-Vazirani algorithm and execution results from real hardware. Using a collection of real and synthetic benchmarks, we show that near-optimal compilation is feasible for systems with small qubit count and limited coherence time. Our results also show that the heuristic method scales to large qubit and gate count, and can efficiently fit programs to execute within the coherence window of the machine.

The main contributions of this paper are as follows:
\begin{itemize}
    \item We develop an end-to-end framework based on constrained optimization to compile high level quantum programs for near term NISQ systems. 
    \item Using real and synthetic benchmarks, we demonstrate that the constraint-based compiler can be used for near-optimal compilation on current and near term systems.
    \item We propose a heuristic method for compiling programs for machines with large qubit and gate counts. The heuristic method uses optimization to fit the execution schedule of a program within the coherence window of the machine. 
    \item   We demonstrate that our heuristic method scales to large programs. For large programs on 128 and 256 qubits, we exhibit cases where the SMT solver can fit all the gates within the allowed coherence window, while a greedy scheduling method cannot.
\end{itemize}

The rest of the paper is organized as follows: Section \ref{sec:related} discusses related work and Section \ref{sec:prelims} provides an overview of NISQ systems and the Scaffold language. Section \ref{sec:overview} presents the key ideas for NISQ compilation. Sections \ref{sec:opt_map_schedule}-\ref{sec:opt_search} develop the near-optimal compilation method using the SMT solver. In Section \ref{sec:heuristic}, we describe a fast heuristic method. Sections \ref{sec:expt_setup}-\ref{sec:results3} present experimental setup and results. 

\section{Related Work}
\label{sec:related}
Many quantum programming languages and compilers have been developed with the goal of simplifying and abstracting quantum programming from the low level details of the hardware. These includes works such as Quipper \cite{quipper1, quipper2}, which is a domain specific language embedded in Haskell, and LIQUi$\ket{}$ \cite{liquid1} which uses the F$\#$ language. These languages offer functionality for quantum circuit description, classical control and compilation and circuit generation. ProjectQ \cite{projectq1, projectq2}, based on Python, is a framework which allows simple quantum circuit description and compilation for different backends. OpenQASM \cite{openqasm1} is a low level language to specify a quantum execution at a gate level. It is used as an interface for near term quantum machines \cite{ibmq}. In this paper, we use the Scaffold language which allows us to describe the quantum circuit at a high level and leverage the rich LLVM compiler infrastructure for automated program analysis and optimization \cite{scaffcc1, llvm}.

In contrast to compilers and frameworks for prior languages, we describe a compilation approach which considers the machine coherence time as a primary constraint. This formally guarantees that the compiled code can finish execution before the hardware qubits decohere. For programs on small NISQ systems, we can also find near optimal compilations, which can help in mitigating errors due to state decay. Our work provides a toolflow which compiles high-level Scaffold programs down to a device-independent intermediate representation using ScaffCC and then efficiently maps and optimizes the intermediate representation for a target device.

Quantum circuit compilation has been studied for different hardware technologies and topologies. Bhattacharjee et al. \cite{ilp1} use an integer linear programming solver to compile small quantum programs for nearest neighbor architectures. However, their approach is applicable only for tiny programs with less than $7$ qubits and $90$ gates. Using the SMT solver based approach, we demonstrate that near optimal compiled code can be synthesized for significantly larger configurations. Guerreschi et al. \cite{intel1} develop a heuristic to schedule quantum circuits on a linear topology and assume that all gates (including swaps) require unit time. Venturelli et al. \cite{ai1} use temporal AI planners for scheduling a certain class of quantum circuits. Heckey et al. \cite{heckey1} develop heuristic compilation techniques for a SIMD gate execution model and assume quantum teleportation based communication. \cite{dousti_mapping, zulehner_mapping, wille_mapping, azim_mapping} are other works on compiling quantum circuits.
In contrast to these approaches, we provide a general end-to-end compilation framework for transforming Scaffold programs to execution ready OpenQASM code. 


Recently, Fu et al. \cite{koen}, developed QuMA, a microarchitecture for QC systems based on superconducting qubits. QuMA takes compiler generated quantum instructions as input and uses micro-instructions to achieve precise timing control of the physical qubits.

\section{Preliminaries}
\label{sec:prelims}

\subsection{NISQ Systems}
NISQ systems encompass near-term quantum computers that are expected to scale up to a few hundred qubits.
They are expected to support a universal gate set, which allows any computation to be expressed in terms of a small number of basis operations or gates. Qubits in these systems have to be isolated from each other, and from the environment, to prevent noise and errors due to unwanted interactions. On the other hand, to perform two qubit (CNOT) gates, certain pairs of qubits should be able to interact strongly without influencing neighboring qubits. Hence, NISQ systems are expected to support limited qubit connectivity where only neighboring qubits can participate in CNOT gates.

Qubits in these systems are expected to have a coherence time of hundreds of microseconds to few milliseconds. The expected gate error rates are in the range $0.001$-$0.01$. These factors imply that only a few thousand operations can be performed before the quantum state decoheres. NISQ systems of this scale can potentially have important applications in quantum chemistry, quantum semidefinite programming, combinatorial optimization and machine learning \cite{nisq}.

This paper focuses on NISQ systems where the qubits are arranged in the form of a grid. We assume nearest neighbor connectivity, where two qubits can participate in a CNOT gate if they are adjacent on the grid. If a CNOT gate is to be performed between two qubits which are not adjacent, the qubits have to be moved to adjacent locations using a series of swap operations. 
 
\subsection{Scaffold: Quantum Programming Language}
Scaffold is a programming language for expressing quantum algorithms. It is an extension of C with quantum types. The user can specify quantum algorithms using a gate set and use familiar C style functions and loops to modularize the code. A useful feature of Scaffold is the ability to specify certain quantum algorithms in classical logic, using {\tt rkqc} modules. These modules allows users to express quantum computation using well known classical logic operations.
A variety of applications have been expressed in Scaffold, we refer the reader to \cite{ali, scaffcc2} for more details.

\subsection{ScaffCC Compiler}
The ScaffCC compiler for Scaffold uses the LLVM compiler infrastructure to compile quantum programs, perform resource estimation and apply error correction. ScaffCC uses the LLVM intermediate representation to transform the program using a set of compilation passes. These passes include transformations such as loop unrolling, procedure cloning, automatic Toffoli and rotation decomposition, and conversion of {\tt rkqc} modules into their quantum equivalents using RevKit \cite{revkit}. Since quantum programs are usually compiled for fixed inputs, ScaffCC includes techniques to resolve classical control dependencies and produce an intermediate output consisting of only quantum gates.
In this paper, we use the ScaffCC compiler as a frontend to extract a gate-level description of the computation, which is then used to synthesize code for NISQ machines.

\section{Optimization Techniques for NISQ Compilation}
\label{sec:overview}
\begin{figure*}
    \centering
    \includegraphics[scale=0.5]{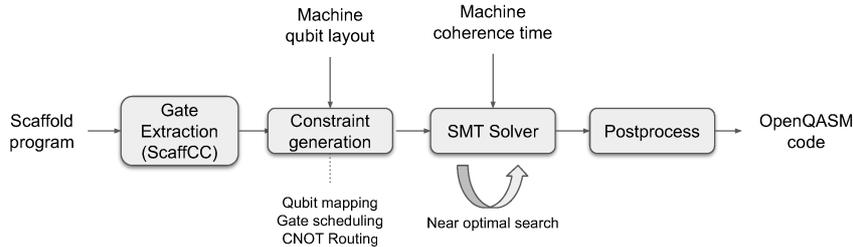}
    \caption{Overview of the compilation process. The compiler extracts gates from the Scaffold program and uses it in conjunction with the machine configuration to solve an SMT optimization problem.}
    \label{fig:overall}
\end{figure*}

Figure \ref{fig:overall} shows our overall compilation approach, accepting  a Scaffold program as input and producing OpenQASM target code. To accomplish this, we first extract gates from the Scaffold program using ScaffCC. The gates extracted from ScaffCC are in terms of program qubits and are independent of the target device. We develop compilation techniques which map this gate representation onto a target device using SMT optimization. In the SMT optimization step, we solve a constrained optimization problem to determine a program mapping, execution schedule and routing decisions which provide an executable with with the minimum makespan (execution time). The makespan is the difference between the finish time of the last gate and the start time of the first gate. Finally, we postprocess the output of the solver, insert routing operations as computed by the SMT optimization and emit OpenQASM code. 

\subsection{Gate Extraction}
The first module inputs a Scaffold program, and uses the ScaffCC compiler \cite{scaffcc1} to extract the LLVM intermediate representation (IR) of the program.
Since NISQ systems are expected to have low coherence time, realistic programs for these systems will only have a small number of gates (hundreds to thousands). This allows us to consider the whole program as a single block for the purpose of optimization. Hence, we unroll all loops and inline all functions in the program to create a single program module. In addition, ScaffCC also performs the rotation and gate decomposition, and classical to quantum module conversion steps, to create a flattened IR for the program.

\begin{figure*}
    \centering
    \begin{subfigure}[b]{0.3\textwidth}
        \centering
        \includegraphics[width=\textwidth]{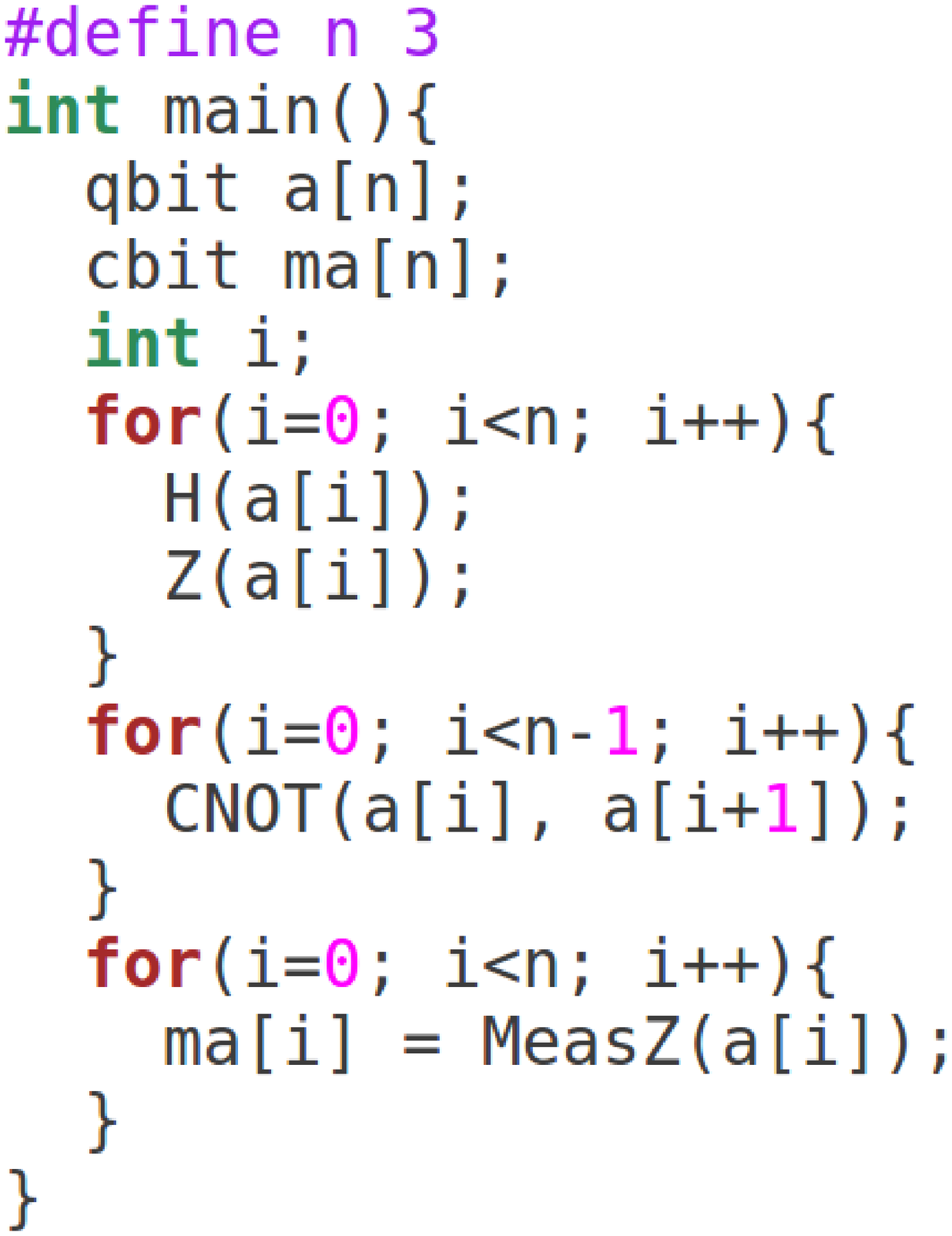}
        \caption{Scaffold program}
        \label{fig:scaffold_program}
    \end{subfigure}
    \quad
    \begin{subfigure}[b]{0.3\textwidth}
        \centering
        \includegraphics[width=\textwidth]{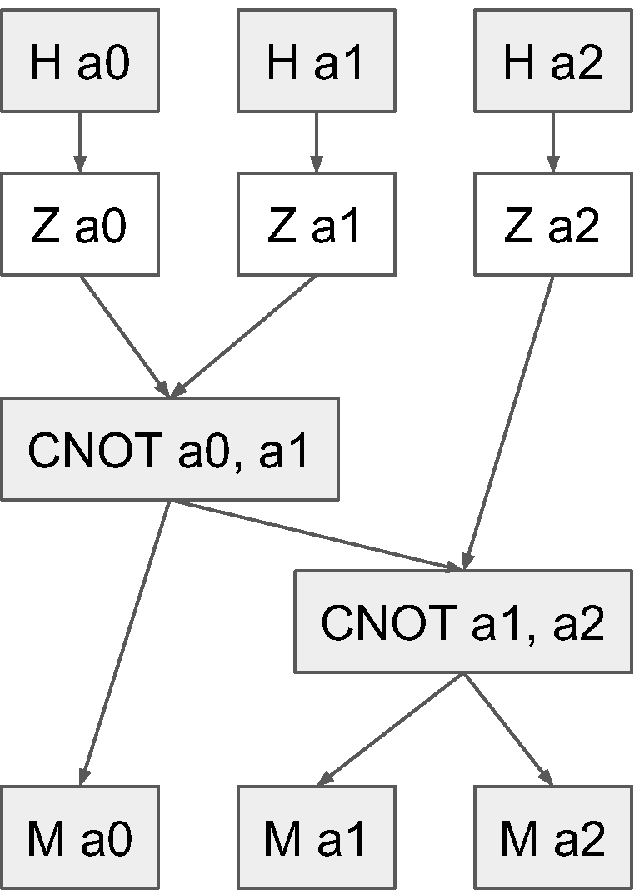}
        \caption{Gate Dependency Graph}
        \label{fig:gate_depend}
    \end{subfigure}
    \quad

\hide{
    \begin{subfigure}[b]{0.25\textwidth}
        \centering
        \includegraphics[width=\textwidth]{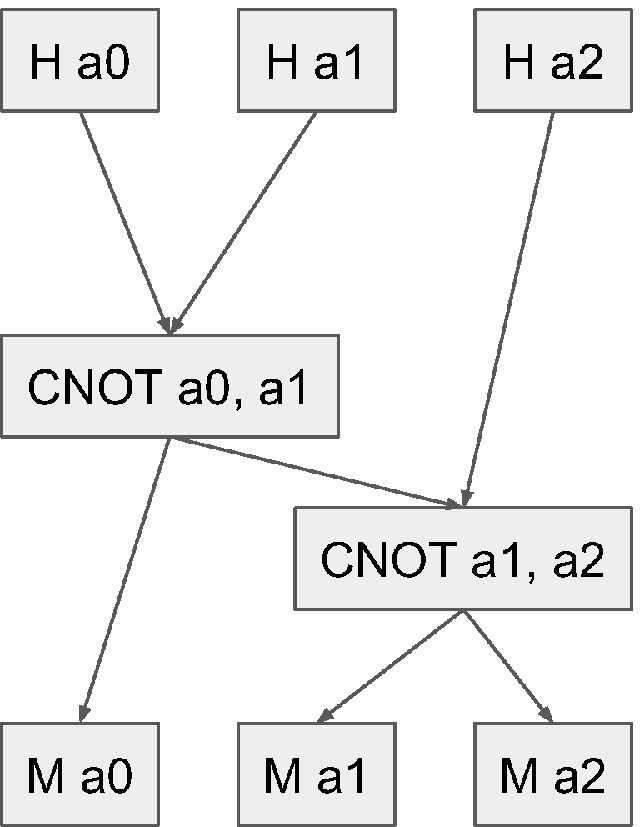}
        \caption{Reduced Dependency Graph}
        \label{fig:gate_reduced}
    \end{subfigure}
}
    \caption{A Scaffold program and the corresponding dependency graph extracted by ScaffCC.}
    \label{fig:gate_extraction}
\end{figure*}

We use the flattened IR to extract gate level information. The gate level information specifies each gate in the program, the qubits it acts on, and its input and output dependencies. The output of this module is summarized as a dependency graph for the program. The vertices of the graph are the gates extracted from the program, and the edges denote the data dependencies between the gates. Each vertex is annotated with the qubits that a gate operates on. For example, Figure \ref{fig:gate_extraction} shows a Scaffold program and its dependency graph.

\subsection{Constraint Generation and Optimization}
The compiler creates an SMT optimization problem which consists of a set of variables to track qubit mappings, and gate start times and durations. The optimization constraints cater to three main factors: 
\begin{enumerate}
    \item \textbf{Qubit Mapping}: The compiler maps program qubits to hardware qubits. The mapping constraint specifies that no two program qubits can be mapped to the same hardware qubit.
    \item \textbf{Gate Scheduling}: For each gate extracted from the Scaffold program, the compiler determines a start time and a duration. For single qubit gates, the duration is the execution time of the operation on the target hardware. For CNOT gates, the control and target qubits may have to be moved using SWAP operations to place the qubits into adjacent locations on the hardware. The execution time for CNOTs includes the duration of the SWAP sequence. 
    Gate scheduling addresses data dependencies through constraints that a gate should start execution only after the gates it depends on finish.
    \item \textbf{CNOT Routing}: To prevent routing conflicts, the compiler uses two routing policies to reason about the SWAP paths of CNOTs. In the first policy, the compiler blocks the rectangle bounded by the control and target qubit, and routes the SWAPs using hardware qubits in the rectangle. In the second policy, the compiler selects one of the two paths along the edge of the bounding rectangle (paths which bend only once) and performs SWAPs along the selected path. In either case, the SMT constraint is that if two CNOT gates overlap in time, their SWAP paths through the hardware qubits should not overlap. 
\end{enumerate}

The SMT optimization simultaneously accounts for all three categories of constraints: qubit mapping decisions affect and are influenced by the gate scheduling and routing decisions. We describe the mapping and scheduling constraints in Section \ref{sec:opt_map_schedule} and the routing constraints in Section \ref{sec:opt_routing}. 

\hide{
Subject to the above constraints, the objective is to minimize the total execution time of the program. As Section \ref{sec:opt_search} describes, rather than seek the true overall minimum, we add a constraint that all gates in the program have to finish execution before a specified time $T$. The compiler uses a binary search procedure to find a near optimal value for $T$. 

The output of the solver is post processed to generate code for the target platform. The post processing phase implements SWAP gates for CNOT routing. It produces target code in the OpenQASM language, which can be directly executed on current quantum hardware. 
}





\section{SMT Optimization: Qubit Mapping and Gate Scheduling}
\label{sec:opt_map_schedule}
In this section, we first describe the setup for the optimization problem, followed by the qubit mapping and gate scheduling constraints.

We process the dependency graph of the program, along with the configuration of the target machine to create a constrained optimization problem. The program is represented as a dependency graph $P=(V,E)$ on $Q$ qubits, where $V$ is the set of gates and $E$ is the set of dependencies. The total number of gates is $G=|V|$. Each dependency in $E$ is a pair of gates $(i,j)$ such that gate $j$ can start only after gate $i$ finishes. We assume that for any qubit, the dependency graph specifies a total ordering of the gates which act on the qubit. Since ScaffCC decomposes multi-qubit gates, each gate in the dependency graph is either a single or two qubit gate.

The machine is represented as an $M$x$N$ grid of hardware qubits. Each qubit is referred to using its location on the 2-D grid. Qubit $(i,j)$ has hardware CNOT connections to qubits $(i+1,j)$ and $(i,j+1)$.
This representation closely models the nearest neighbhor connections in real systems such as the IBM 16-qubit system \cite{ibmq} and the system in development at Google \cite{coupling3, googleb}. In this paper, we apply swap operations for communication in a restoring manner i.e., if we apply a set of swaps to change the qubit ordering before a CNOT, we apply the same swaps after the CNOT to restore the qubit order.

\subsection{Qubit Mapping}
A program qubit $i$ is mapped to a hardware qubit $(q_x[i], q_y[i])$. We add the following constraints to ensure that mappings respect the distinctness constraint:
\begin{gather}
q_x[i] \in [1,M],  \forall i \in [1,Q] \label{cons:map1}\\
q_y[i] \in [1,N],  \forall i \in [1,Q] \label{cons:map2}\\
q_x[i] \ne q_x[j] \lor q_y[i] \ne q_y[j], \forall i,j \in [1,Q] \text{s.t.}  i < j \label{cons:map3}
\end{gather}

\subsection{Gate Scheduling}
For every gate $j$, the solver should determine a start time $t[j]$ and duration $d[j]$. The finish time of a gate is $t[j] + d[j]$. First, we constrain the start and finish times to lie within the machine's coherence threshold ($T$):

\begin{gather}
t[j] \in [1,T], \forall j \in [1,G] \\
t[j] + d[j] \le T, \forall j \in [1,G]
\end{gather}

For any single qubit gate $j$, we can set the duration variable using the duration of the corresponding hardware gate i.e., $d[j] == \tau(\text{type}(j))$. Here, $\tau$ is a mapping which specifies the duration for each gate type. For example, for any Hadamard gate $j$, we can set the duration as 1 time slot by hard wiring $d[j] == 1$. 

For CNOT gates, the optimizer has to account for the duration of the hardware CNOT, and the time required to move the qubits in place before and after the hardware CNOT. 
If the $L_1$ distance between the control and target qubit is $l$, the time taken for the CNOT is the sum of the durations of $l-1$ SWAP gates, the hardware CNOT, and the restoring sequence of $l-1$ SWAP gates. We express this duration as:

\begin{gather}
\text{define} |x| = \text{If-Then-Else}(x\ge 0, x, -x)\\
\text{dist}(c, t) = |q_x[c] - q_x[t]| + |q_y[c] - q_y[t]|\\
d[j] = 2(\text{dist}(\text{ctrl}(j),\text{targ}(j))-1)*\tau(SWAP) + \tau(CNOT) \label{cons:cnot_duration}
\end{gather}
We add constraint \ref{cons:cnot_duration} for every gate $j$ which is of type CNOT. We note that the time required for a SWAP can be halved by implementing a meet in the middle policy where both control and target qubits move in parallel. However, it increases the number of parallel operations among nearby qubits and can potentially cause more crosstalk errors. In this paper, we assume that only the control qubit moves to the target qubit using a series of swaps.

Finally, we can represent gate dependencies, by enforcing that a gate $j$ can start only after its dependent gate $i$ has finished: 
\begin{gather}
    t[j] \ge t[i] + d[i], \forall (i,j) \in E \label{cons:dependancy}
\end{gather}
    
\section{SMT Optimization: CNOT Routing}
\label{sec:opt_routing}
CNOTs which occur between program qubits which are at non-adjacent locations require communication using SWAP gates. In this section, we describe two communication routing policies and a pruning strategy to reduce the number of routing constraints.

\begin{figure*}[h]
    \centering
    \begin{subfigure}[b]{0.3\textwidth}
        \centering
        \includegraphics[width=\textwidth]{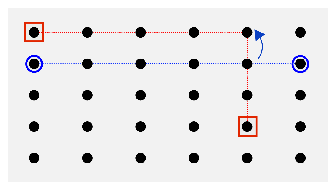}
        \caption{Intersecting swap paths}
        \label{fig:intersect_swap}
    \end{subfigure}
    \quad
    \begin{subfigure}[b]{0.3\textwidth}
        \centering
        \includegraphics[width=\textwidth]{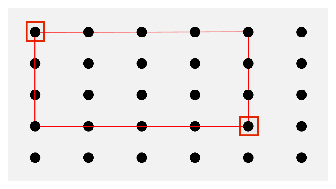}
        \caption{Rectangle Reservation}
        \label{fig:rr}
    \end{subfigure}
    \quad
    \begin{subfigure}[b]{0.3\textwidth}
        \centering
        \includegraphics[width=\textwidth]{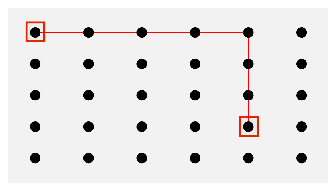}
        \caption{One bend paths}
        \label{fig:1bp}
    \end{subfigure}
    
    \caption{CNOT Routing. Figure \ref{fig:intersect_swap} illustrates the need for non-overlapping swap paths. Figures \ref{fig:rr} and \ref{fig:1bp} illustrate the two routing policies used in our compiler.}
    \label{fig:routing_policies}
\end{figure*}

We observe that the swap paths taken by concurrent CNOTs should not intersect. In Figure \ref{fig:intersect_swap}, we illustrate the necessity of having spatially non-overlapping swap paths. If the control qubits corresponding to the red and blue CNOT pairs are moving towards their respective targets, it is possible that the control qubits can swap with each other and get deviated from their routing  path. In such cases, the length of the path from the control to the target is no longer the $L_1$ distance, and it is difficult to quantify the CNOT duration exactly in constraints  \ref{cons:cnot_duration} and \ref{cons:dependancy}. If we consider the case where swap paths overlap spatially, but qubits use distinct hardware edges at any given time, then the overall qubit mapping can depend on the relative order in which the swap sequences are executed (not illustrated). 

These scenarios motivate us to spatially restrict the swap paths of CNOTs which overlap in time. We use two routing policies: rectangle reservation and 1-bend paths. These policies are inspired from similar policies in VLSI routing \cite{gonzalez}. We first explain the two routing policies and then discuss a pruning strategy which reduces the number of routing constraints required for compilation. 

\subsection{Rectangle reservation} In rectangle reservation, we reserve the 2-D region bounded by the control and target qubit locations for the duration of the CNOT. For example, in Figure \ref{fig:rr}, the highlighted rectangle is reserved for the duration of the CNOT. For this policy, we define solver variables and constraints to check if two CNOT rectangles overlap. The SMT constraint is that if two CNOTs overlap in time, their bounding rectangles should not overlap in space. The solver reserves the rectangle for the duration of the CNOT, and the exact swap path within the rectangle is computed during post-processing. 

To implement rectangle reservation, we add variables which track the top-left and bottom-right locations of each  CNOT in the program. Consider a CNOT gate $i$. We can define the top-left corner using variables $(l_x[i], l_y[i])$ and the bottom-right corner using variables $(r_x[i], r_y[i])$. These variables can be defined using min and max relationships on the control and target qubit locations. Denote the control-target rectangle of a CNOT $i$ as $R_i$. Using these variables the following constraint detects whether the rectangles of two CNOTs $i$ and $j$ overlap in space and time:
\begin{gather}
OverlapInSpace(R_i,R_j) = \lnot(l_x[i] > r_x[j] \lor  r_x[i] < l_x[j] \lor l_y[i] > r_y[j] \lor  r_y[i] < l_y[j]) \\  
OverlapInTime(i,j) = \lnot(t[i] > t[j] + d[j] \lor t[j] > t[i] + d[i])
\end{gather}
The routing constraint for any pair of CNOTs $i$ and $j$ is 
\begin{gather}
OverlapInTime(i,j) \implies \lnot OverlapInSpace(i,j)
\end{gather}

\subsection{1-Bend Paths} For the second routing policy, we restrict CNOTs to routing paths which bend at most once on the 2-D grid. There are two such paths along the edges of the bounding rectangle of the control and target qubit. This policy is very similar to dimension ordered routing.

For example, in Figure \ref{fig:1bp}, the swaps can be routed using the highlighted red path along the top edge of the bounding rectangle.
In this case, we require the solver to pick one of the two paths using a variable which determines the bend point or routing junction. 1-bend paths are advantageous because they block less resources than rectangle reservation at run time. However, the solver requires additional compile time to determine the exact path during optimization. 

For 1-bend paths, we can write constraints similar to rectangle reservation to check overlap in space. For a CNOT $i$, the solver uses two junction variables $b_x[i]$ and $b_y[i]$ to determine the location of the bend point. The two segments of the path are the control to junction segment, and the junction to target segment. We can consider these segments as rectangles and apply the overlap check as in rectangle reservation. For a CNOT $i$, denote the control to junction segment as $R^{cb}_i$ and the junction to target segment as $R^{bt}_i$. The spatial overlap condition for two CNOTs $i$ and $j$ is:
\begin{align}
Overlap(i,j) = & OverlapInSpace(R^{cb}_i, R^{cb}_j) \lor OverlapInSpace(R^{cb}_i, R^{bt}_j)  \lor \\ \notag
& OverlapInSpace(R^{bt}_i, R^{cb}_j)  \lor  OverlapInSpace(R^{bt}_i, R^{bt}_j) 
\end{align}
As in rectangle reservation, we impose a condition that the paths should not overlap in space if the gates overlap in time.

\hide{
In rectangle reservation the solver does not determine the exact path used for routing. It blocks the rectangle and leaves the path computation for post-processing. For 1-bend paths the solver selects one of the two paths during compile time. Hence, rectangle reservation is easier than 1-bend paths at  compile time. However, rectangle reservation can block too many resources at run time and increase the program execution time.
}

\subsection{Transitive Closure based Pruning}
\begin{figure}
    \centering
    \includegraphics[scale=0.5]{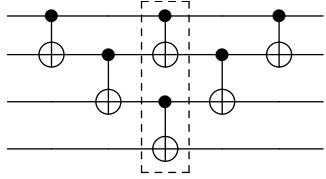}
    \caption{A circuit to illustrate transitive closure based pruning of routing constraints. In this circuit, only the two CNOTs in the dashed box need a routing constraint. None of the other pairs of CNOTs can overlap in time.}
    \label{fig:tc}
\end{figure}

Evaluating routing constraints during SMT optimization is computationally expensive because these constraints have more literals than the mapping and scheduling constraints. 
We observe that we do not require routing constraints for every pair of program CNOTs. For any CNOT gate, any gate which depends directly or indirectly on the gate cannot overlap with it. Similarly, any gate on which the CNOT depends cannot overlap with it. These overlaps are avoided by the gate dependency constraint (constraint \ref{cons:dependancy}). For example, in Figure \ref{fig:tc}, only the two CNOTs in the highlighted box can overlap in time. We can determine whether two CNOTs can overlap by computing the transitive closure of the dependency graph. For any node in the graph, the transitive closure gives us the set of ancestors and descendants in the dependency order. Any CNOT gate which is not an ancestor or descendant can potentially overlap with the CNOT. For every pair of overlapping gates determined using the transitive closure algorithm, we add a routing constraint. We can compute the transitive clousure efficiently using the Floyd-Warshall algorithm \cite{clrs}. In a perfectly sequential program, the transitive closure pruning allows us to avoid routing constraints entirely. On our benchmarks, we found that transitive closure based pruning can provide up to $20$x reduction in the number of routing constraints.

\hide{
\subsubsection{Crosstalk constraint} In current quantum systems gate operations on a qubit influence the state of neighboring qubits which are not participating in the operations. These crosstalk errors stem from undesired qubit interactions are known to occur for both single and two qubit gates \cite{coupling2, coupling1}. On some experimental architectures a controlled Z gate between a pair of gates blocks any gate on the neighboring qubits \cite{coupling3}. We incorporate partial crosstalk considerations in our compiler by introducing crosstalk awareness for CNOT gates. When a CNOT gate is required between two qubits, we protect the region around the swap path and the actual hardware CNOT. We can easily modify our routing constraints to block a larger area around the CNOT path. 

For rectangle reservation, this translates to an expanded rectangle around the control and target qubit. For 1-bend paths, we transform the paths into small rectangles: one rectangle from the control qubit to the bend junction and another rectangle from the junction to the target. To protect qubits within distance $D$ of the CNOT, we modify the $(l_x, l_y)$ and $(r_x, r_y)$ variables to be $(l_x-D, l_y-D)$ and $(r_x+D, r_y+D)$ and use the same constraints as before. SImilarly, we can modify the variables associated with  of the control to junction and junction to target paths to transform them into rectangles.
}

\section{OPT Algorithm: Near Optimal Search}
\label{sec:opt_search}
The objective of the solver is to minimize the total execution time or makespan of the schedule. We introduce a dummy gate $G+1$, which depends on every gate in the program: 
\begin{gather}
    t[G+1] \ge t[i] + d[i], \forall i \in [1,G] 
\end{gather}
The optimization objective is to minimize the start time of the dummy gate:
\begin{gather}
    \text{minimize} \quad t[G+1]
\end{gather}

We can minimize this objective function using the Optimization Modulo Theory (OMT) solver in Z3 \cite{omt_z3}. To compute a qubit mapping and gate schedule which minimizes the execution time, we set up an optimization problem using the mapping, scheduling and routing constraints discussed earlier. The qubit mapping and gate start time variables interlink the three sets of constraints and the objective. We specify this optimization problem using the Z3 APIs and the solver finds the optimal solution. 

In our experiments we found that the Z3 OMT solver is quite slow in practice because it searches for the exact optimal solution. To use the satisfiability checker in Z3, we rewrite the objective function as a constraint, $t[G+1] \le T_{max}$. We can search for a good value of $T_{max}$ using a binary search procedure. We start with an estimate of the upper bound $U=T$ (the coherence
window of the machine) and a lower bound $L=0$. In every step of the binary search, we maintain the invariant that the optimization is satisfiable for $T_{max}=U$ and unsatisfiable for $T_{max}=L$. Since the optimal value is guarenteed to lie within $(L, U]$, we compute the approximation quality as $\eta = U/(L+1)$. We terminate the binary search when the $\eta < 1+\epsilon$, where $\epsilon$ is a small constant. In our experiments we denote this search procedure as the OPT algorithm. We set $\epsilon = 0.1$ to obtain a solution where the execution duration is at most 1.1x factor more than the optimum.



\section{Heuristic Method}
\label{sec:heuristic}
In our experiments, we found that the near optimal solution can be computed for circuits with small qubit and gate count, which is characteristic of programs on current and short term quantum systems. For future systems with larger qubit counts and coherence time, we design a fast heuristic compilation method.

The primary scalability bottleneck for the solver is that it performs qubit mapping, gate scheduling and routing simultaneously in an exponentially large search space. We design an optimization based heuristic which obtains a fast qubit mapping and uses the solver to schedule and route operations. This approach preserves the flexibility offered by the optimization problem and obtains solutions which are reasonably close to the optimum.

We separate the compilation problem into two phases: in the first phase, we map qubits to hardware and in the second phase, we schedule and route gates. To find a good mapping, we employ a greedy strategy which minimizes the total number of SWAP operations. The intuition behind the greedy mapping is as follows: for every pair of qubits in the program, we compute a weight $w$, as the number of CNOTs between the pair. If a pair has a large number of CNOTs (higher weight), the qubits should be mapped close together in the hardware to reduce the amount of communication. Consider a mapping $\pi: Q\mapsto H$, where $Q$ is the set of program qubits and $H$ is the set of hardware qubits. For two program qubits $q_i$ and $q_j$, we denote the weight as $w_{ij}$. The total number of swaps required to perform CNOTs between $q_i$ and $q_j$ is $d(\pi(q_i), \pi(q_j))$, where $d$ is a distance function which accurately models the hardware topology. Therefore, the objective is to:
\begin{gather}
\displaystyle{\minimize_{\pi} \sum_{i,j \in Q}w_{ij}d(\pi(q_i), \pi(q_j)) }
\end{gather}

Since it is NP-hard to optimize this function, we use a greedy strategy to obtain a mapping. We denote the weight of a qubit as the total number of CNOTs it participates in. We consider qubits in non-increasing order of weight. To map a new qubit to the hardware, we find the location which minimizes its sum of weighted distances to already mapped qubits. 

\hide{
\RestyleAlgo{boxruled}
\begin{algorithm}
 \KwData{$Q$: set of program qubits, $H$: set of hardware qubits, $d$: distance function based on hardware topology, $w_{ij}$: CNOT frequencies for each pair of program qubits)}
 \KwResult{$\pi:Q\mapsto H$ }
 $W_i = \sum_{j\in Q}w_{ij}$\;
 $\bar{Q}$: sorted $Q$ in non-increasing order of $W$\;
 $\pi = \phi$\;
 $P = \phi$ \tcp*{set of mapped program qubits}
 $A = H$ \tcp*{set of available hardware qubits}
 \For{$q_i \in \bar{Q}$}{
    $h^* = \argmin_{h \in A}$ $\sum_{j \in P} w_{ij}d(h, \pi(q_j))$ \;
    $\pi(q_i) = h^*$ \tcp*{map program qubit $q_i$ to $h^*$} 
    $P = P \cup {i}$\;
    $A = A \setminus {h^*}$\;
 }
 \caption{Greedy qubit mapping. In each iteration of the for loop, the heaviest remaining vertex ($q_i$) is mapped to the best available hardware qubit ($h^*$).}
 \label{algo:greedy_map}
\end{algorithm}
}

After computing the greedy mapping, we can perform gate scheduling and routing. We perform this in two stages: greedy scheduling and routing, followed by refinement using the SMT solver if necessary. 

To find a greedy execution schedule, we iteratively schedule the earliest gate which is ready i.e., a gate whose dependent gates have finished execution. For rectangle reservation, we can incorporate routing into this algorithm, by computing the earliest time at which the ready gate can be scheduled without conflicting with previously scheduled gates. Similarly, for one bend paths, we greedily select the bend point which gives the ready gate the earliest start time. If the execution duration of the greedy schedule fits within the coherence window of the machine, we use the computed mapping and execution schedule. If the length of the greedy schedule exceeds the coherence window of the machine, we use the SMT solver to search for a refined execution schedule. The greedy mapping is used as input to an optimization formulation where we have only scheduling and routing constraints. In other words, we omit constraints \ref{cons:map1}-\ref{cons:map3} and hard wire the mapping variables to the greedy mapping. Then, we add a constraint that the execution duration is less than the coherence window of the machine and search for a satisfiable solution. This approach ensures that the solution produced by the solver respects the coherence time of the machine and the routing constraints. 

\section{Experimental Setup}
\label{sec:expt_setup}

\paragraph{Quantum Machine}
We assume a 2-D grid of qubits with nearest neighbor connectivity. The gate durations used in our experiments are listed in Table \ref{tab:gate_time}. The grid sizes used for our experiments are listed in Table \ref{tab:config}. 
For real experiments, we use the IBM 16-qubit machine using the IBM Quantum Experience APIs \cite{ibmq}.

\begin{figure}[t]
    \centering
    \includegraphics[scale=0.6]{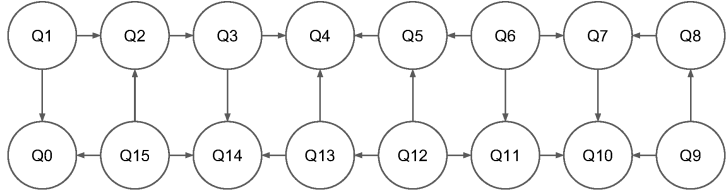}
    \caption{Qubit layout in the 16-qubit IBM machine {\tt IBMQ16 Rueschlikon}. The CNOTs in this machine, shown by arrows, are uni-directional.}
\label{fig:ibmqx5}
\end{figure}

\paragraph{Implementation for Real Hardware}
The layout of the IBM 16-qubit {\tt IBMQ16 Rueschlikon} system is shown in Figure \ref{fig:ibmqx5} \cite{ibmq}. All hardware CNOTs in this system are uni-directional. The coherence time of the machine is  $~100$ microseconds. We normalize all times using the time for a single control pulse ($80$ns). We use the gate durations listed in Table \ref{tab:gate_time}. We use well known transformations to implement SWAP gates using CNOTs, and reversed CNOT gates using Hadamard gates \cite{Mermin}. We generate output code in the OpenQASM language \cite{openqasm1}. 


\begin{table}[t]
\centering
\footnotesize
\begin{minipage}[b]{0.4\linewidth}
\centering
\begin{tabular}{|l|l|}
\hline 
Gate           & \shortstack{Duration \\ (timeslots)} \\ \hline \hline
CNOT           & 8        \\
Measure        & 5        \\
X              & 2        \\
Y              & 2        \\
H              & 1        \\ \hline
Z              & 0        \\
S, S$^\dagger$ & 0        \\
T, T$^\dagger$ & 0        \\ \hline
SWAP           & 24       \\ \hline
\end{tabular}
\caption{Gate durations. 
}
\label{tab:gate_time}
\end{minipage}
\quad
\begin{minipage}[b]{0.4\linewidth}
\centering
\begin{tabular}{|l|l|l|} \hline 
Qubits & M  & N  \\ \hline \hline
8      & 2  & 4  \\ \hline
16     & 2  & 8  \\ \hline
32     & 4  & 8  \\ \hline
64     & 8  & 8  \\ \hline
128    & 8  & 16 \\ \hline
256    & 16 & 16 \\ \hline                     
\end{tabular}
\caption{Machine configurations.}
\label{tab:config}
\end{minipage}
\end{table}

\paragraph{Benchmark}
We present results using real and synthetic programs. Our real benchmark consists of a set of circuits for the Bernstein-Vazirani (BV) algorithm \cite{Mermin, bernsteinvazirani}, Ising model \cite{ising}, and Square Root using Grover's search \cite{ali,grover}. These benchmarks are implemented in Scaffold\footnote{Ising model and Square Root are available at \url{https://github.com/epiqc/ScaffCC}}. The synthetic benchmark consists of programs where we apply uniformly chosen random gates from the set $\{$CNOT, H, X, Y, Z, T, S, T$^\dagger$, S$^\dagger\}$ on randomly chosen qubits. These programs are generated without information about the machine topology. The gate set used in our experiments is: . 

\paragraph{Algorithms}
We study two methods: the near optimal search method where the solver simultaneously performs mapping, scheduling and routing (Section \ref{sec:opt_search}) and the heuristic method where we use greedy qubit mapping (Section \ref{sec:heuristic}). We refer to the first method as OPT. The approximation threshold $\epsilon$ for OPT is set to be $0.1$.

\paragraph{Metrics} We compare the algorithms on compilation time and execution time. The execution time or makespan of the generated schedule is the difference of the finish time of the last gate and the start time of the first gate. For the BV algorithm, we also report the correctness of the algorithm as measured on the IBM 16-qubit system.

\paragraph{Implementation} Our framework implements pre and post-processing steps in Python3.5 and the core solver routines in C++. We use the C++ interface to the Z3 SMT solver 4.6.0 to construct and solve the optimization problem. Our compilation runs are performed on an Intel Xeon machine (3.20GHz, 128GB main memory).

\hide{
\paragraph{Experiments} We first present compilation and execution results for the BV algorithm on {\tt IBMQ16 Rueschlikon}. Then, we study the performance of the OPT algorithm on the synthetic benchmark. We discuss the scalability bottleneck of the OPT algorithm and the effect of routing policies. Next, we study the heuristic schedules on real and synthetic benchmarks and evaluate the scalability of the heuristic algorithm. 
}

\section{Bernstein-Vazirani Algorithm on Real Hardware}
\label{sec:results1}
We present real results from compiling and executing programs for the BV algorithm on the 16-qubit IBM hardware. These experiments use the OPT compiler.

\begin{figure}[t]
    \centering
    \includegraphics[scale=0.5]{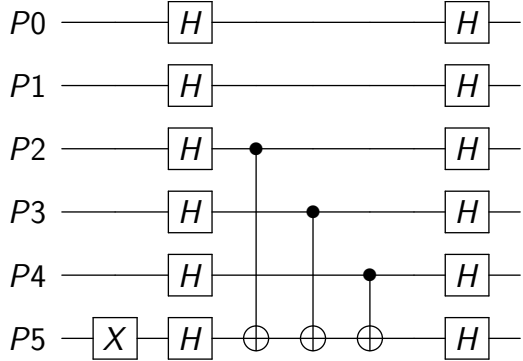}
    \caption{Bernstein-Vazirani Algorithm with 6 qubits for a hidden bitstring ``00111''. We denote this configuration as (6,3). All qubits are initialized to the zero state.}
    \label{fig:bv63}
\end{figure}

Given a function  
$f(x):\{0,1\}^n \rightarrow \{0,1\}$ of the form
$\mathbf{a}\cdot\mathbf{x} \pmod{2}$ where 
$\mathbf{a}\in\{0,1\}^n$ is an unknown bitstring, the BV algorithm computes the $n$ bits of $\mathbf{a}$ using a single query to a quantum implementation of the function. In contrast, a classical algorithm will require at least $n$ queries to extract all the bits of $\mathbf{a}$. To achieve this, the algorithm first puts all the qubits in a superposition state and passes them through an oracle implementation of the function. Using a quantum effect called phase kickback, it can efficiently recover the hidden bits. In Figure \ref{fig:bv63}, we show a quantum circuit for 6 bits $P0-P5$. This circuit implements the oracle corresponding to the hidden bit string ``00111''. When this circuit is executed on a machine, and the qubits are measured, the hidden string is expected as output.

\begin{figure}[t]
    \centering
   \begin{subfigure}[b]{0.7\textwidth}
        \centering
    \includegraphics[width=1\linewidth]{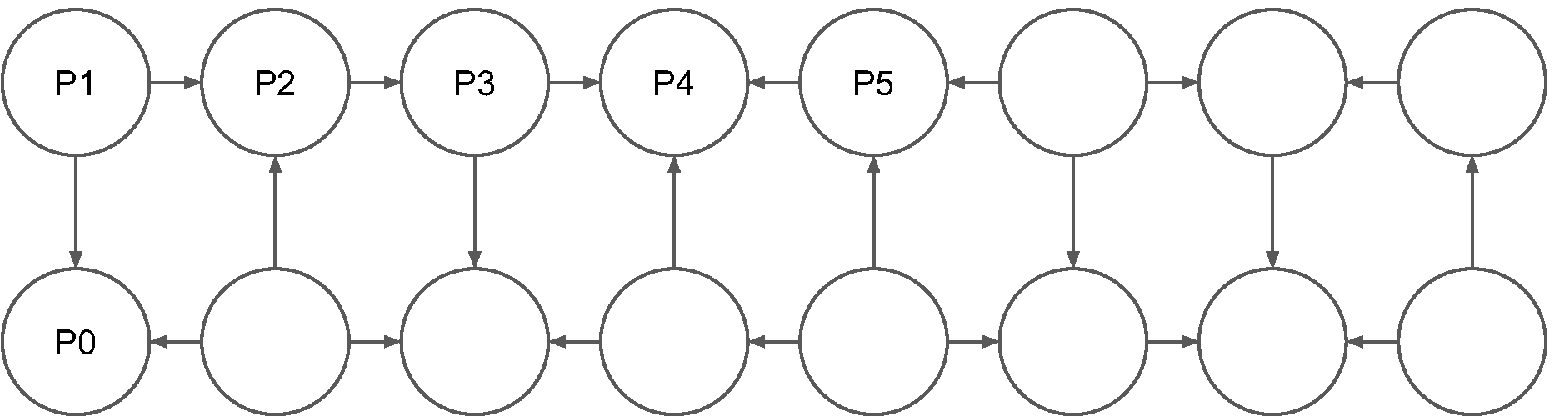}
        \caption{Naive mapping of program qubits based on their id}
        \label{fig:bv63mapping_prog}
    \end{subfigure}
    \quad    
   \begin{subfigure}[b]{0.7\textwidth}
        \centering
    \includegraphics[width=1\linewidth]{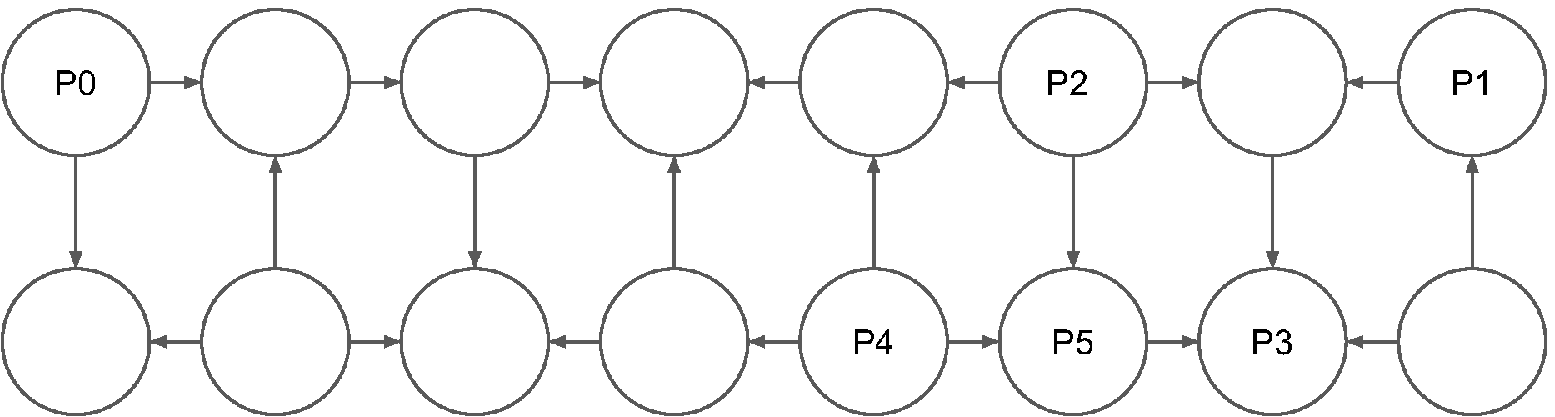}
        \caption{Mapping computed by the \opt algorithm}
        \label{fig:bv63mapping_opt}
    \end{subfigure}

    \caption{An illustration of qubit mapping for the Bernstein-Vazirani $(6,3)$ program. The mapping produced by our compiler accounts for CNOT communication and places the communicating qubits in adjacent locations. P0 and P1 have no communication, and therefore can be distant. A naive program order mapping of the qubits can be suboptimal for communication and overall execution time.}
    \label{fig:bv63mapping}
\end{figure}

If we map the program qubits to the hardware qubits based on their id, we will obtain the mapping shown in Figure \ref{fig:bv63mapping_prog}. We can see that qubits $P2$ and $P5$ will have to use SWAP gates to perform the required CNOT. In Figure \ref{fig:bv63mapping_opt}, we illustrate the mapping obtained by the OPT algorithm. We can see that the compiler places qubit $P5$ on a degree 3 node in the system to minimize the distances to the control qubits $P2$, $P3$ and $P4$, which are placed in adjacent locations. It places qubits $P0$ and $P1$ at arbitrary locations because they do not communicate with other qubits. 

We created BV programs in Scaffold, with 4-8 qubits and hidden bitstrings of varying length. The circuit parameters are shown in Table \ref{tab:bv_results}.  For each program, the compiler generates OpenQASM code which can be executed using the IBM Quantum Experience APIs. We performed experiments on the IBM QASM simulator and the 16-qubit machine, and used $8192$ trials. Each trial corresponds to one execution of the program. A trial is a success if the measured classical output matches the hidden bitstring. We note that the experiments on the simulator and the real hardware use the same OpenQASM code. This allows us to verify the correctness of the synthesized code in the absence of noise and gate errors on the real hardware. 

\begin{table}[t]
\centering
\footnotesize
\begin{tabular}{|l|l|l|l|l|}
\hline
Qubits & \shortstack{Hidden \\ String} & \shortstack{Compile \\ Time (s)} & \shortstack{Success \\ \shortstack{rate \\{\tt IBMQ16 Rueschlikon}}} & \shortstack{Success \\ \shortstack{rate \\ {\tt sim}}} \\ \hline \hline
4               & 001                    &  1                          & 0.42                  & 1 \\ \hline
4               & 111                    &  1                         & 0.32                  & 1 \\ \hline
6               & 00001                  &  1                         & 0.70                  & 1\\ \hline
6               & 00111                  &  1                         & 0.26                  & 1 \\ \hline
6               & 11111                  &  1                         & $<$0.1                  & 1 \\ \hline
8               & 0000001                &  1                         & 0.54                  & 1 \\ \hline
8               & 0000111                &  1                         & 0.43                  & 1\\ \hline
8            & 0011111                & 2                           & $<$0.1   & 1 \\ \hline
\end{tabular}
\caption{Results from executions of Bernstein-Vazirani Algorithm on {\tt IBMQ16 Rueschlikon} and the IBM QASM Simulator {\tt sim}. The success rates on the machine and the simulator demonstrate correct compilation and execution.}
\label{tab:bv_results}
\end{table}

\begin{figure}[t]
    \centering
    \includegraphics[scale=0.5]{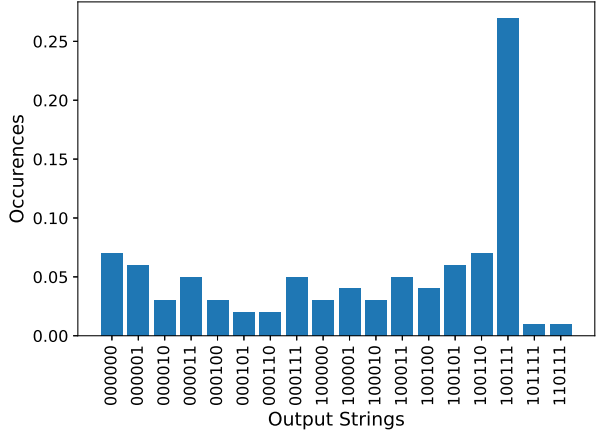}
    \caption{Top 20 outcomes from executions of the Bernstein-Vazirani program for the hidden bitstring ``00111'' (See Figure \ref{fig:bv63}). The program qubits are measured in the order P5, P0, P1, P2, P3, P4 since P5 is expected to be 1. The required bitstring dominates the output distribution.} 
    \label{fig:bv63_qx5}
\end{figure}

From Table  \ref{tab:bv_results}, we can see that the OPT algorithm finds a near optimal schedule for these programs in a few seconds. We report the success rates observed on the simulator and the real machine as the ratio of successful trials to the total number of trials. We can see that the success rate on the simulator is perfect, validating the correctness of the compilation pipeline. On the real machine, we see reasonably high success probability in all but two cases. For the (6,3) program we show the measured output distribution in Figure \ref{fig:bv63_qx5}. We see that the required bitstring dominates the output distribution. We also observe the effect of single and two qubit errors which corrupt the output and produce strings with one or more bits flipped. 

\section{Evaluation of the OPT Algorithm}
\label{sec:results2}
\subsection{Compilation Time}
In this experiment we use the synthetic benchmark to study the compilation time of the OPT algorithm.
The random benchmarks used in our experiments are shown in Table \ref{tab:opt_results}. For each program, we run the OPT algorithm with rectangle reservation and 1-bend path policies to obtain a $1.1$x approximation of the makespan. We report the compilation time and makespan of the schedule. In three cases we see that the solver times out (24 hours) while trying to refine the schedule using binary search. For these cases, we report the makespan of the best solution obtained.

From Table \ref{tab:opt_results}, we can see that, for circuits with a small number of qubits and gates, the solver provides near optimal compilation. Increasing the number of qubits or gates, increases compilation time. We can understand this trend using Table \ref{tab:solver_behavior}, where we categorize programs according to qubit and gate count. For programs with small qubit and gate count, finding the near optimal solution is feasible because the search space is small. For large qubit count, the search space of mappings becomes exponentially high. For large gate count, the cost of evaluating mappings by scheduling and routing gates, becomes prohibitively high.

However, for current systems and near term systems (5-32 qubits) with limited coherence time, we can use the OPT algorithm to obtain the best compilation, instead of relying on heuristics. In particular, for the 16-qubit IBM system, the coherence time is 1250 timeslots. From the table, we can see that, for programs with 8 and 16 qubits which have optimal makespan less than 1250, OPT can compute the best schedule quickly.


\subsection{Effect of Routing Policy}
From Table \ref{tab:opt_results}, we can see that both routing policies obtain the same makespan, except in two cases where the makespan for the 1-bend path policy is better than rectangle reservation by $10.5\%$. The 1-bend path policy increases the compile time, by up to $3.3$x factor, because the solver has to determine the exact swap path for each CNOT using additional decision variables. For programs with small number of gates, the OPT algorithm finds qubit mappings where no swapping is required. Hence, we do not require routing in such cases and choice of routing policy does not matter. For programs with large number of gates relative to the qubit count (8 qubits and 256 or 512 gates), the routing policy becomes important because more pairs of program qubits perform CNOTs. In these cases, it is beneficial to use 1-bend paths. We expect that the benefits of 1-bend paths will be more when the circuits have higher parallelism and when qubits perform CNOTs with many other qubits. In such scenarios, finding non-overlapping 1-bend paths will be beneficial compared to blocking large parts of the machine using rectangle reservation.

\begin{table}[t]
\centering
\footnotesize
\begin{tabular}{|l|l|l|l|l|l|}
\hline
\multicolumn{2}{|c|}{\shortstack{Circuit \\ Properties}} & \multicolumn{2}{c|}{\shortstack{Makespan \\ (timeslots)}} & \multicolumn{2}{c|}{\shortstack{Compile \\ Time (s)}} \\ \hline \hline
Qubits          & Gates         & RR       & 1BP       & RR          & 1BP          \\ \hline 
8                        & 64                     & 29                & 29                 & 0                    & 0                     \\ \hline
8                        & 128                    & 351               & 351                & 20                   & 31                    \\ \hline
8                        & 256                    & 742               & 664                & 270                  & 298                   \\ \hline
8                        & 512                    & 1484              & 1328               & 2729                 & 2479                  \\ \hline
16                       & 64                     & 31                & 31                 & 1                    & 3                     \\ \hline
16                       & 128                    & 175               & 175                & 60                   & 162                   \\ \hline
16                       & 256                    & 234               & 234                & 418                  & 949                   \\ \hline
16                       & 512                    & 5000              &5000                    & \multicolumn{2}{|c|}{timeout}                                     \\ \hline
32                       & 64                     & 19                & 19                 & 3                    & 6                     \\ \hline
32                       & 128                    & 92                & 92                 & 598                  & 1210                  \\ \hline
32                       & 256                    & 53                & 53                 & 253                  & 845                   \\ \hline
32                       & 512                    & 8806                  &8806                    &  \multicolumn{2}{|c|}{timeout}                                     \\ \hline
64                       & 64                     & 16                & 16                 & 8                    & 15                    \\ \hline
64                       & 128                    & 29                & 29                 & 31                   & 73                    \\ \hline
64                       & 256                    & 39                & 39                 & 1283                 & 2349                  \\ \hline
64                       & 512                    & 10000                   & 10000                   &  \multicolumn{2}{|c|}{timeout}                                     \\ \hline
\end{tabular}

\caption{Compilation time and makespan for the synthetic benchmark using the OPT algorithm. We compare two routing policies: rectangle reservation (RR) and 1-bend paths (1BP). }
\label{tab:opt_results}
\end{table}

\begin{table}[t]
\centering
\footnotesize
\begin{tabular}{cccc}
                       &                           & \multicolumn{2}{c}{\textsc{Qubits}}                                                                                                                                                                                          \\
                       &                           & \multicolumn{1}{c}{Low}                                                                             & \multicolumn{1}{c}{High}                                                                                      \\ \cline{3-4} 
\multirow{2}{*}{\rotatebox{90}{\textsc{Gates}}} & \multicolumn{1}{c|}{Low}  & \multicolumn{1}{c|}{\begin{tabular}[c]{@{}c@{}}Near optimum \\ is feasible\end{tabular}}               & \multicolumn{1}{c|}{\begin{tabular}[c]{@{}c@{}}Many mappings:  $H!/(H-Q)!$ 
\end{tabular}}                      \\ \cline{3-4} 
                       & \multicolumn{1}{c|}{High} & \multicolumn{1}{c|}{\begin{tabular}[c]{@{}c@{}}Large time per mapping: $O(T^G)$  \end{tabular}} & \multicolumn{1}{c|}{\begin{tabular}[c]{@{}c@{}}Many mappings +\\ large time per mapping \end{tabular}} \\ \cline{3-4} 
\end{tabular}
\caption{Solver runtime behavior for programs with different qubit and gate count. $H$: number of machine qubits, $Q$: number of program qubits, $T$: coherence time of the machine, $G$: number of gates in the program.}
\label{tab:solver_behavior}
\end{table}

\section{Evaluation of the Heuristic Algorithm}
\label{sec:results3}
\subsection{Comparison of Optimal and Heuristic Schedules}
In this section, we study the schedules obtained using the heuristic algorithm. Recall that the heuristic algorithm uses a greedy strategy for mapping program qubits to hardware qubits. This method aims to reduce the total number of swaps in the synthesized code. Once a mapping is computed, a greedy schedule is computed and the SMT solver is invoked to fit the schedule to the coherence window of the machine. 

Table \ref{tab:greedy_results} evaluates the heuristic schedules for the synthetic benchmark. To compare the heuristic and optimal mappings  independent of assumed machine coherence times and without introducing any inefficiency in gate scheduling or routing, we modify the heuristic method so that the SMT solver searches for a near-optimal schedule for the greedy mapping. In other words, we specify a large bound (100000) for the coherence window and obtain the best gate execution schedule possible using the two algorithms. We measure the loss factor due to the heuristic as the ratio of the makespan of the heuristic schedule to the schedule produced by the OPT algorithm. 

Comparing Table \ref{tab:opt_results} and \ref{tab:greedy_results}, we can see that the makespans of schedules computed by the heuristic algorithm are longer than OPT.   The loss factor of the heuristic algorithm is $1.6$x and $1.4$x (geomean) for rectangle reservation and 1-bend path policies, respectively. In all cases, the heuristic computes a schedule within $2$ minutes. 

For three cases (starred), the heuristic finds solutions better than the solutions computed by OPT. These are cases where the OPT algorithm timed out while searching for the near optimal solution. In the worst case, we see that the makespan of the heuristic schedule is 17x higher than OPT on one program. In this program, OPT computes a solution which requires no swap operations, resulting in low makespan. 

\begin{table}[t]
\centering
\footnotesize
\begin{tabular}{|l|l|l|l|l|l|l|l|}
\hline
\multicolumn{2}{|c|}{\begin{tabular}[c]{@{}c@{}}Circuit\\ Properties\end{tabular}} & \multicolumn{2}{l|}{\begin{tabular}[c]{@{}c@{}}Makespan\\ (timeslots)\end{tabular}} & \multicolumn{2}{l|}{\begin{tabular}[c]{@{}c@{}}Makespan\\ Loss factor\end{tabular}} & \multicolumn{2}{l|}{\begin{tabular}[c]{@{}c@{}}Compile \\ Time (s)\end{tabular}} \\ \hline \hline
Qubits              & Gates              & RR            & 1BP           & RR                                       & 1BP                                      & RR                                       & 1BP                                       \\ \hline
8                   & 64                 & 29            & 29            & 1.00                                     & 1.00                                     & 1                                        & 0                                         \\ \hline
8                   & 128                & 351           & 351           & 1.00                                     & 1.00                                     & 0                                        & 1                                         \\ \hline
8                   & 256                & 858           & 742           & 1.16                                     & 1.12                                     & 1                                        & 1                                         \\ \hline
8                   & 512                & 2031          & 1718          & 1.37                                     & 1.29                                     & 10                                       & 2                                         \\ \hline
16                  & 64                 & 30            & 31            & 0.97                                     & 1.00                                     & 0                                        & 0                                         \\ \hline
16                  & 128                & 341           & 234           & 1.95                                     & 1.34                                     & 0                                        & 1                                         \\ \hline
16                  & 256                & 798           & 859           & 3.41                                     & 3.67                                     & 0                                        & 2                                         \\ \hline
16                  & 512$^*$                & 3750          & 2810          & 0.75                                     & 0.56                                     & 98                                       & 50                                        \\ \hline
32                  & 64                 & 71            & 72            & 3.74                                     & 3.79                                     & 1                                        & 0                                         \\ \hline
32                  & 128                & 585           & 486           & 6.36                                     & 5.28                                     & 0                                        & 1                                         \\ \hline
32                  & 256                & 922           & 584           & 17.40                                 & 11.02                                    & 1                                          & 1                                         \\ \hline
32                  & 512$^*$                & 2812          & 2030          & 0.32                                     & 0.23                                     & 77                                       & 19                                        \\ \hline
64                  & 64                 & 16            & 16            & 1.00                                     & 1.00                                     & 0                                        & 0                                         \\ \hline
64                  & 128                & 79            & 82            & 2.72                                     & 2.83                                     & 0                                        & 1                                         \\ \hline
64                  & 256                & 146           & 146           & 3.74                                     & 3.74                                     & 1                                        & 2                                         \\ \hline
64                  & 512$^*$               & 2031          & 1484          & 0.20                                         & 0.15                                          & 1                                        & 20                                        \\ \hline
\end{tabular}
\caption{Compilation time and makespan for the synthetic benchmark using the heuristic algorithm. We report the loss factor of the heuristic as the ratio of makespan of the heuristic schedule to the optimal schedule from Table \ref{tab:opt_results}.}
\label{tab:greedy_results}
\end{table}

\subsection{Evaluation on Real Benchmarks}
Next, Table \ref{tab:heur_real} evaluates the heuristic algorithm on the Ising model and Square root benchmarks on a 128-qubit (8x16) grid. We can see that, for the square root program with 78 qubits and 1515 gates, the compiler requires only 75 seconds of compilation time. Across programs and routing policies, the maximum compilation time is less than 5 minutes. We see no significant difference in makespan for rectangle reservation and 1-bend paths. This is because these benchmarks are highly sequential and do not have a lot of overlapping CNOTs. In contrast, on synthetic benchmarks which have more parallelism, we see (from Table \ref{tab:greedy_results}) that 1-bend paths can provide up to $1.4$x improvement in execution duration compared to rectangle reservation.

\begin{table}[t]
\centering
\footnotesize
\begin{tabular}{|l|l|l|l|l|l|l|}
\hline
\multicolumn{3}{|c|}{Benchmark}    & \multicolumn{2}{c|}{Makespan (timeslots)} & \multicolumn{2}{c|}{\begin{tabular}[c]{@{}c@{}}Compile Time (s)\end{tabular}} \\ \hline \hline
Name              & Qubits & Gates & RR            & 1BP           & RR                                       & 1BP                                      \\ \hline 
Ising model 1      & 5      & 668   & 279           & 279           & 1                                        & 1                                        \\ \hline
Ising model 2      & 10     & 1513  & 288           & 288           & 1                                        & 1                                        \\ \hline
Square root (n=3) & 17     & 244   & 7479          & 7343          & 1                                        & 3                                        \\ \hline
Square root (n=4) & 30     & 502   & 27463         & 26991         & 61                                       & 74                                       \\ \hline
Square root (n=5) & 47     & 843   & 63531         & 61250         & 3                                        & 154                                      \\ \hline
Square root (n=6) & 78     & 1515  & 160000        & 160000        & 75                                       & 262                                      \\ \hline
\end{tabular}

\caption{Evaluation of the heuristic algorithm on two real benchmarks. We can see that all programs are compiled within 5 minutes.}
\label{tab:heur_real}
\end{table}

\subsection{Scalability of the Heuristic Algorithm}

\begin{table}[t]
\centering
\footnotesize
\begin{tabular}{|l|l|l|l|l|l|l|}
\hline
Qubits                                                                & 8   & 16   & 32   & 64   & 128   & 256   \\ \hline
\begin{tabular}[c]{@{}l@{}}Coherence Time (ms)\end{tabular}         & 50  & 100  & 200  & 400  & 800   & 1600  \\ \hline
\begin{tabular}[c]{@{}l@{}}Coherence  Time (timeslots)\end{tabular} & 625 & 1250 & 2500 & 5000 & 10000 & 20000 \\
\hline
\end{tabular}

\caption{Coherence times for our scalability experiment. These coherence times were obtained by scaling the coherence time of {\tt IBMQ16 Rueschlikon}.}
\label{tab:coh_scale}
\end{table}

In this experiment, we use the heuristic algorithm and configurations for near term machines to determine whether the SMT solver can compile programs to fit within the coherence window of the machine. We created a benchmark with $8$ to $256$ qubits with depth $2$ to $10$. For a program with qubit count $q$ and depth $d$, the number of gates generated is $qd$. The coherence times for this experiment are shown in Table \ref{tab:coh_scale}. These times are obtained by scaling the  coherence time of the IBM 16 qubit machine ($100$us) by 2x, for every 2x increase in machine size.

\begin{figure}[t]
    \centering
    \includegraphics[scale=0.5]{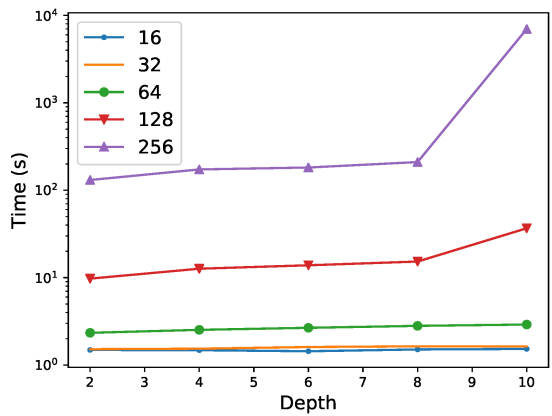}
    \caption{Compilation time of the heuristic algorithm for input programs with different qubit count and depth. Each line represent a particular qubit count. For (128 qubits, depth 10) and (256 qubits, depth 10), the output of the heuristic fits within the allowed coherence window, whereas a greedy method overshoots the window.}
    \label{fig:scaling}
\end{figure}

For each program in this benchmark, we use the coherence time for the machine with the same qubit count, and compile it using the heuristic algorithm. We plot the compilation times for different qubit counts and program depths in Figure \ref{fig:scaling}. In all cases, we found that the heuristic method was able to find a feasible schedule where all gates fit within the specified coherence window. For programs with less than 128 qubits, we can find a feasible execution schedule within 100 seconds. 

In two cases, 128 qubits with 1280 gates, and 256 qubits with 2560 gates, we found that the SMT optimization is crucial to fit the execution schedule within the coherence window. For the 128 qubit case, the greedy schedule (earliest ready gate first schedule) required 10898 timesteps, which is higher than the allowed coherence threshold. The SMT solver was able to optimize the schedule to fit it within the coherence window, and produced a schedule which executes in 9999 timesteps. We observed similar behavior for the case with 256 qubits. In general, when the program's makespan is comparable to the machine's coherence threshold, heuristics may not be effective. In such cases, the SMT solver based approach is particularly useful to carefully arrange the gates within the available coherence window.

\section{Conclusions}
In this paper, we developed a compiler for Scaffold, a high level language, targeted for near term quantum systems with hundreds of qubits. We developed a constraint based compilation method which uses an SMT solver to simultaneously map program qubits to hardware qubits, schedule and route gates, while minimizing total execution time. Using real and synthetic benchmarks, we showed that it is feasible to obtain near optimal compilations for current and near term NISQ machines. For larger programs and machine sizes, we developed a heuristic method which uses optimization to fit the program to the coherence window of the machine. We demonstrated that this method is scalable, and succeeds in finding coherence compliant schedules.

\section{Acknowledgements}
This work is funded in part by EPiQC, an NSF Expedition in
Computing, under grants CCF-1730449/1730082, in part by NSF PHY-1818914 and a research gift from Intel.



\bibliographystyle{elsarticle-num}


\begin{thebibliography}{10}
\expandafter\ifx\csname url\endcsname\relax
  \def\url#1{\texttt{#1}}\fi
\expandafter\ifx\csname urlprefix\endcsname\relax\def\urlprefix{URL }\fi
\expandafter\ifx\csname href\endcsname\relax
  \def\href#1#2{#2} \def\path#1{#1}\fi

\bibitem{nmr1}
L.~M.~K. Vandersypen, I.~L. Chuang, {NMR techniques for quantum control and
  computation}, Rev. Mod. Phys. 76 (2005) 1037--1069.
\newblock \href {http://dx.doi.org/10.1103/RevModPhys.76.1037}
  {\path{doi:10.1103/RevModPhys.76.1037}}.

\bibitem{nmr2}
J.~A. Jones, M.~Mosca, R.~H. Hansen, Implementation of a quantum search
  algorithm on a quantum computer, Nature 393.

\bibitem{trappedion1}
J.~I. Cirac, P.~Zoller, {Quantum Computations with Cold Trapped Ions}, Phys.
  Rev. Lett. 74 (1995) 4091--4094.
\newblock \href {http://dx.doi.org/10.1103/PhysRevLett.74.4091}
  {\path{doi:10.1103/PhysRevLett.74.4091}}.

\bibitem{trappedion2}
T.~P. Harty, D.~T.~C. Allcock, C.~J. Ballance, L.~Guidoni, H.~A. Janacek, N.~M.
  Linke, D.~N. Stacey, D.~M. Lucas, {High-Fidelity Preparation, Gates, Memory,
  and Readout of a Trapped-Ion Quantum Bit}, Phys. Rev. Lett. 113 (2014)
  220501.
\newblock \href {http://dx.doi.org/10.1103/PhysRevLett.113.220501}
  {\path{doi:10.1103/PhysRevLett.113.220501}}.

\bibitem{superc1}
C.~Rigetti, J.~M. Gambetta, S.~Poletto, B.~L.~T. Plourde, J.~M. Chow, A.~D.
  C\'orcoles, J.~A. Smolin, S.~T. Merkel, J.~R. Rozen, G.~A. Keefe, M.~B.
  Rothwell, M.~B. Ketchen, M.~Steffen, Superconducting qubit in a waveguide
  cavity with a coherence time approaching 0.1 ms, Phys. Rev. B 86 (2012)
  100506.
\newblock \href {http://dx.doi.org/10.1103/PhysRevB.86.100506}
  {\path{doi:10.1103/PhysRevB.86.100506}}.

\bibitem{superc2}
J.~Majer, J.~M. Chow, J.~M. Gambetta, J.~Koch, B.~R. Johnson, J.~A. Schreier,
  L.~Frunzio, D.~I. Schuster, A.~A. Houck, A.~Wallraff, A.~Blais, M.~H.
  Devoret, S.~M. Girvin, R.~J. Schoelkopf, Coupling superconducting qubits via
  a cavity bus, Nature 449.

\bibitem{nisq}
J.~Preskill, {Quantum Computing in the NISQ era and beyond} (2018).
\newblock \href {http://arxiv.org/abs/arXiv:1801.00862}
  {\path{arXiv:arXiv:1801.00862}}.

\bibitem{survey1}
M.~H. Devoret, R.~J. Schoelkopf, {Superconducting Circuits for Quantum
  Information: An Outlook}, Science 339~(6124) (2013) 1169--1174.
\newblock \href {http://dx.doi.org/10.1126/science.1231930}
  {\path{doi:10.1126/science.1231930}}.

\bibitem{Devitt2013}
S.~J. Devitt, A.~M. Stephens, W.~J. Munro, K.~Nemoto, Requirements for
  fault-tolerant factoring on an atom-optics quantum computer, Nature
  Communications 4 (2013) 2524 EP --, article.

\bibitem{ali}
A.~Javadi-Abhari, {Towards a Scalable Software Stack for Resource Estimation
  and Optimization in General-Purpose Quantum Computers}, {PhD} dissertation,
  Princeton University (2017).

\bibitem{z3}
L.~de~Moura, N.~Bj{\o}rner, {Z3: An Efficient SMT Solver}, in: C.~R.
  Ramakrishnan, J.~Rehof (Eds.), Tools and Algorithms for the Construction and
  Analysis of Systems, Springer Berlin Heidelberg, Berlin, Heidelberg, 2008,
  pp. 337--340.

\bibitem{quipper1}
A.~S. Green, P.~L. Lumsdaine, N.~J. Ross, P.~Selinger, B.~Valiron, {Quipper: A
  Scalable Quantum Programming Language}, in: Proceedings of the 34th ACM
  SIGPLAN Conference on Programming Language Design and Implementation, PLDI
  '13, ACM, New York, NY, USA, 2013, pp. 333--342.
\newblock \href {http://dx.doi.org/10.1145/2491956.2462177}
  {\path{doi:10.1145/2491956.2462177}}.

\bibitem{quipper2}
A.~S. Green, P.~L. Lumsdaine, N.~J. Ross, P.~Selinger, B.~Valiron, Quipper: A
  scalable quantum programming language, SIGPLAN Not. 48~(6) (2013) 333--342.
\newblock \href {http://dx.doi.org/10.1145/2499370.2462177}
  {\path{doi:10.1145/2499370.2462177}}.

\bibitem{liquid1}
D.~Wecker, K.~M. Svore, {LIQU$|>$: A Software Design Architecture and
  Domain-Specific Language for Quantum Computing} (2014).
\newblock \href {http://arxiv.org/abs/arXiv:1402.4467}
  {\path{arXiv:arXiv:1402.4467}}.

\bibitem{projectq1}
D.~S. Steiger, T.~Haner, M.~Troyer, {ProjectQ: An Open Source Software
  Framework for Quantum Computing}, Quantum 2, 49 (2018)\href
  {http://arxiv.org/abs/arXiv:1612.08091} {\path{arXiv:arXiv:1612.08091}},
  \href {http://dx.doi.org/10.22331/q-2018-01-31-49}
  {\path{doi:10.22331/q-2018-01-31-49}}.

\bibitem{projectq2}
{Project Q}, \url{https://projectq.ch/}, accessed: 2018-05-16.

\bibitem{openqasm1}
A.~W. Cross, L.~S. Bishop, J.~A. Smolin, J.~M. Gambetta, {Open Quantum Assembly
  Language} (2017).
\newblock \href {http://arxiv.org/abs/arXiv:1707.03429}
  {\path{arXiv:arXiv:1707.03429}}.

\bibitem{ibmq}
{IBM Quantum Experience},
  \url{https://quantumexperience.ng.bluemix.net/qx/devices}, accessed:
  2018-05-16.

\bibitem{scaffcc1}
A.~Javadi-Abhari, S.~Patil, D.~Kudrow, J.~Heckey, A.~Lvov, F.~T. Chong,
  M.~Martonosi, {ScaffCC: A Framework for Compilation and Analysis of Quantum
  Computing Programs}, in: Proceedings of the 11th ACM Conference on Computing
  Frontiers, CF '14, ACM, New York, NY, USA, 2014, pp. 1:1--1:10.
\newblock \href {http://dx.doi.org/10.1145/2597917.2597939}
  {\path{doi:10.1145/2597917.2597939}}.

\bibitem{llvm}
C.~Lattner, V.~Adve, {LLVM: A Compilation Framework for Lifelong Program
  Analysis \& Transformation}, in: Proceedings of the International Symposium
  on Code Generation and Optimization: Feedback-directed and Runtime
  Optimization, CGO '04, IEEE Computer Society, Washington, DC, USA, 2004, pp.
  75--.

\bibitem{ilp1}
D.~Bhattacharjee, A.~Chattopadhyay, Depth-optimal quantum circuit placement for
  arbitrary topologies (2017).
\newblock \href {http://arxiv.org/abs/arXiv:1703.08540}
  {\path{arXiv:arXiv:1703.08540}}.

\bibitem{intel1}
G.~G. Guerreschi, J.~Park, Two-step approach to scheduling quantum circuits
  (2017).
\newblock \href {http://arxiv.org/abs/arXiv:1708.00023}
  {\path{arXiv:arXiv:1708.00023}}.

\bibitem{ai1}
D.~Venturelli, M.~Do, E.~Rieffel, J.~Frank, Compiling quantum circuits to
  realistic hardware architectures using temporal planners, 2017 Quantum Sci.
  Technol.\href {http://arxiv.org/abs/arXiv:1705.08927}
  {\path{arXiv:arXiv:1705.08927}}, \href
  {http://dx.doi.org/10.1088/2058-9565/aaa331}
  {\path{doi:10.1088/2058-9565/aaa331}}.

\bibitem{heckey1}
J.~Heckey, S.~Patil, A.~JavadiAbhari, A.~Holmes, D.~Kudrow, K.~R. Brown,
  D.~Franklin, F.~T. Chong, M.~Martonosi, {Compiler Management of Communication
  and Parallelism for Quantum Computation}, in: Proceedings of the Twentieth
  International Conference on Architectural Support for Programming Languages
  and Operating Systems, ASPLOS '15, ACM, New York, NY, USA, 2015, pp.
  445--456.
\newblock \href {http://dx.doi.org/10.1145/2694344.2694357}
  {\path{doi:10.1145/2694344.2694357}}.

\bibitem{dousti_mapping}
M.~J. Dousti, M.~Pedram,
  \href{http://dl.acm.org/citation.cfm?id=2492708.2492917}{{Minimizing the
  Latency of Quantum Circuits During Mapping to the Ion-trap Circuit Fabric}},
  in: Proceedings of the Conference on Design, Automation and Test in Europe,
  DATE '12, EDA Consortium, San Jose, CA, USA, 2012, pp. 840--843.
\newline\urlprefix\url{http://dl.acm.org/citation.cfm?id=2492708.2492917}

\bibitem{zulehner_mapping}
A.~Zulehner, A.~Paler, R.~Wille, {An Efficient Methodology for Mapping Quantum
  Circuits to the IBM QX Architectures} (2017).
\newblock \href {http://arxiv.org/abs/arXiv:1712.04722}
  {\path{arXiv:arXiv:1712.04722}}.

\bibitem{wille_mapping}
R.~Wille, O.~Kesz{\"o}cze, M.~Walter, P.~Rohrs, A.~Chattopadhyay, R.~Drechsler,
  {Look-ahead schemes for nearest neighbor optimization of 1D and 2D quantum
  circuits}, 2016 21st Asia and South Pacific Design Automation Conference
  (ASP-DAC) (2016) 292--297.

\bibitem{azim_mapping}
A.~Farghadan, N.~Mohammadzadeh,
  \href{https://onlinelibrary.wiley.com/doi/abs/10.1002/cta.2335}{{Quantum
  circuit physical design flow for 2D nearest-neighbor architectures}},
  International Journal of Circuit Theory and Applications 45~(7) (2017)
  989--1000.
\newblock \href
  {http://arxiv.org/abs/https://onlinelibrary.wiley.com/doi/pdf/10.1002/cta.2335}
  {\path{arXiv:https://onlinelibrary.wiley.com/doi/pdf/10.1002/cta.2335}},
  \href {http://dx.doi.org/10.1002/cta.2335} {\path{doi:10.1002/cta.2335}}.
\newline\urlprefix\url{https://onlinelibrary.wiley.com/doi/abs/10.1002/cta.2335}

\bibitem{koen}
X.~Fu, M.~A. Rol, C.~C. Bultink, J.~van Someren, N.~Khammassi, I.~Ashraf,
  R.~F.~L. Vermeulen, J.~C. de~Sterke, W.~J. Vlothuizen, R.~N. Schouten, C.~G.
  Almudever, L.~DiCarlo, K.~Bertels, {An Experimental Microarchitecture for a
  Superconducting Quantum Processor}, in: Proceedings of the 50th Annual
  IEEE/ACM International Symposium on Microarchitecture, MICRO-50 '17, ACM, New
  York, NY, USA, 2017, pp. 813--825.
\newblock \href {http://dx.doi.org/10.1145/3123939.3123952}
  {\path{doi:10.1145/3123939.3123952}}.

\bibitem{scaffcc2}
{ScaffCC Compiler}, \url{https://github.com/epiqc/ScaffCC}, accessed:
  2018-05-16.

\bibitem{revkit}
M.~Soeken, S.~Frehse, R.~Wille, R.~Drechsler, {RevKit}: An open source toolkit
  for the design of reversible circuits, in: Reversible Computation 2011, Vol.
  7165 of Lecture Notes in Computer Science, 2012, pp. 64--76, {RevKit is
  available at www.revkit.org}.

\bibitem{coupling3}
A.~Fowler, \href{https://meetings.aps.org/Meeting/MAR18/Session/V33.9}{Towards
  sufficiently fast quantum error correction}.
\newline\urlprefix\url{https://meetings.aps.org/Meeting/MAR18/Session/V33.9}

\bibitem{googleb}
{Google Bristlecone System},
  \url{https://ai.googleblog.com/2018/03/a-preview-of-bristlecone-googles-new.html},
  accessed: 2018-05-16.

\bibitem{gonzalez}
T.~F. Gonzalez, D.~Serena, Complexity of pairwise shortest path routing in the
  grid, Theoretical Computer Science 326~(1) (2004) 155 -- 185.
\newblock \href {http://dx.doi.org/https://doi.org/10.1016/j.tcs.2004.06.027}
  {\path{doi:https://doi.org/10.1016/j.tcs.2004.06.027}}.

\bibitem{clrs}
T.~H. Cormen, C.~E. Leiserson, R.~L. Rivest, C.~Stein, Introduction to
  Algorithms, Third Edition, 3rd Edition, The MIT Press, 2009.

\bibitem{omt_z3}
N.~Bj{\o}rner, A.-D. Phan, L.~Fleckenstein, {$\nu$Z - An Optimizing SMT
  Solver}, in: C.~Baier, C.~Tinelli (Eds.), Tools and Algorithms for the
  Construction and Analysis of Systems, Springer Berlin Heidelberg, Berlin,
  Heidelberg, 2015, pp. 194--199.

\bibitem{Mermin}
N.~D. Mermin, Quantum Computer Science: An Introduction, Cambridge University
  Press, New York, NY, USA, 2007.

\bibitem{bernsteinvazirani}
E.~Bernstein, U.~Vazirani, Quantum complexity theory, in: Proceedings of the
  Twenty-fifth Annual ACM Symposium on Theory of Computing, STOC '93, ACM, New
  York, NY, USA, 1993, pp. 11--20.
\newblock \href {http://dx.doi.org/10.1145/167088.167097}
  {\path{doi:10.1145/167088.167097}}.

\bibitem{ising}
R.~Barends, A.~Shabani, L.~Lamata, J.~Kelly, A.~Mezzacapo, U.~L. Heras,
  R.~Babbush, A.~G. Fowler, B.~Campbell, Y.~Chen, Z.~Chen, B.~Chiaro,
  A.~Dunsworth, E.~Jeffrey, E.~Lucero, A.~Megrant, J.~Y. Mutus, M.~Neeley,
  C.~Neill, P.~J.~J. O'Malley, C.~Quintana, P.~Roushan, D.~Sank,
  A.~Vainsencher, J.~Wenner, T.~C. White, E.~Solano, H.~Neven, J.~M. Martinis,
  Digitized adiabatic quantum computing with a superconducting circuit, Nature
  534 (2016) 222 EP --.

\bibitem{grover}
L.~K. Grover, {A Fast Quantum Mechanical Algorithm for Database Search}, in:
  Proceedings of the Twenty-eighth Annual ACM Symposium on Theory of Computing,
  STOC '96, ACM, New York, NY, USA, 1996, pp. 212--219.
\newblock \href {http://dx.doi.org/10.1145/237814.237866}
  {\path{doi:10.1145/237814.237866}}.

\end{thebibliography}

\end{document}